\documentstyle[preprint,aps,prb,tighten]{revtex}
\def\be{\begin{equation}}
\def\ee{\end{equation}}
\def\bea{\begin{eqnarray}}
\def\eea{\end{eqnarray}}
\def\bmat{\left(\begin{array}}
\def\emat{\end{array}\right)}

\begin{document}
\preprint
\widetext
\title{A Brownian Motion Model of Parametric Correlations in
Ballistic Cavities}
\author{
A. M. S. Mac\^edo
}
\address{
Departamento de F\'{\i}sica, Universidade Federal de Pernambuco\\
50670-901 Recife, PE, Brazil\\}
\date{\today}
\maketitle
\widetext
\begin{abstract}
A Brownian motion model is proposed to study parametric correlations in the
transmission eigenvalues of open ballistic cavities. We find interesting
universal properties when the eigenvalues are rescaled at the hard edge of the
spectrum.
We derive a formula for the power spectrum of the fluctuations
of transport observables as a response to an external adiabatic perturbation.
Our formula correctly recovers the Lorentzian-squared behaviour obtained
by semiclassical approaches for the correlation function of conductance
fluctuations.
\par
\noindent
Pacs numbers: 72.10.Bg, 05.60.+w, 05.40.+j, 05.45.+b
\end{abstract}
\narrowtext
\
\section{Introduction}
The theory of random matrices has been applied to many branches of physics,
such as complex nuclei \cite{5c1}, disordered metallic grains
\cite{5c2,1c15,2c32}, the general theory of quantum chaotic systems \cite{5c3},
 disordered conductors \cite{2c30,2c18,2c23,amsm1} , random surfaces
\cite{5c4}, QCD \cite{5c5} and more recently strongly correlated many-particle
systems \cite{5c6,5c6.1}. One of the most striking and robust predictions of
 random matrix theory (RMT), which has far reaching implications in all these
different fields is the celebrated Wigner-Dyson statistics \cite{5c7,3c11},
which states that the probability distribution of level spacings (usually
 regarded as energy eigenvalues of some random Hamiltonian) is a universal
function if the eigenvalues are measured in units of the mean level spacing.
 In the case of quantum chaotic systems, it is believed that the success of RMT
depends on the chaotic dynamics of
 the equivalent classical system.
This dependence seems to be so strong that many results derived from random
matrix ensembles, such as the Wigner-Dyson statistics, have been used as
signatures of quantum chaos \cite{5c3}.
\par
An interesting extension
\cite{5c8,5c8.1,5c9,5c10,5c11,5c12,5c13,5c13.1,5c14,5c15,5c16,5c17,5c18,5c19,5c20} of RMT is to consider variations in the energy levels
of the physical system resulting from external adiabatic perturbations. These
perturbations could be, for instance, a change in the Aharonov-Bohm flux
through a ring, or a slight alteration in the position of some impurity in a
 disordered metallic grain, or a variation in the strength of an external
applied magnetic field or even simply a tiny modification in the geometry of
the sample.
The theory, subsequently developed by Simons, Altshuler, Lee and others
\cite{5c14,5c15,5c16,5c17,5c18,5c19,5c20}
  to account for this effect, has proven to be a significant extension of RMT,
which we shall call parametric
 random matrix theory (PRMT).  Remarkably, it has been demonstrated that PRMT,
while incorporating all features of RMT, predicts new universal statistics
after appropriate rescaling of the physical parameters, and thus provides an
 even more powerful characterization of quantum chaotic dynamics. PRMT contains
only two system dependent parameters: the mean level spacing, $\Delta$, and the
mean square gradient of levels, $C_0$, which is defined as the ensemble average
\be
C_0=\left\langle \left( {\partial \varepsilon_i(U) \over \partial U}
\right)^2 \right\rangle,
\ee
where $\varepsilon_i\equiv E_i/\Delta$ denote renormalized energy levels and
$U$
is the parameter that controls the strength of the external perturbation. A
striking prediction of PRMT is that the $n$-point correlation function for
density of states fluctuations becomes a  universal expression if the energy
 levels are measured in units of $\Delta$ and the perturbation parameter, $U$,
is renormalized by $C_0$. This universality has been derived analytically for
disordered metallic grains with orthogonal, unitary and symplectic symmetries
\cite{5c15,5c16,5c17} and has been verified numerically in a number of chaotic
systems \cite{5c6,5c6.1,5c15,5c21,5c22,5c23,5c24,5c25}.
\par
We remark that the universality obtained in the framework of PRMT, and
similarly that of Wigner-Dyson statistics, applies only for levels away from
the edge of the support of the spectrum. This statement can, in principle, be
formally justified by means of renormalization group arguments. A simple
renormalization group procedure has
 recently been devised by Br\'ezin and Zinn-Justin \cite{5c26}. These authors
have shown that there are two kinds of fixed points in the renormalization
group equations: a stable gaussian one governing the behavior of the system at
 the bulk of the spectrum and an unstable one governing a small crossover
region around the endpoint of the support
 where the average density goes to zero as a power law. The attractive gaussian
fixed point supports the general
 validity of the Wigner-Dyson statistics everywhere at the bulk of the
spectrum. It is therefore natural to expect that the results of PRMT are of
similar general validity, although an explicit proof is not yet available.
\par
At the edge of the spectrum, however, Wigner-Dyson statistics breaks down and
an entirely new regime of universal
 statistics emerges. There are two kinds of edges in RMT: a hard edge and a
soft edge \cite{5c27,5c28,5c29,5c30}, where the eigenvalue support is bounded
and unbounded respectively. The Gaussian ensembles have two soft edges,
 while the Laguerre ensembles exhibit a hard and a soft edge. There are also
ensembles with two hard edges, like for instance the Jacobi ensembles.
\par
The hard edge of the Laguerre ensemble is important in the description of many
physical systems such as disordered
 metallic conductors, ballistic cavities and QCD \cite{5c5}. In a recent paper
Slevin and Nagao \cite{5c31} have introduced a group of matrices, which they
called $\Omega$-matrices, that gives rise to a very powerful way of studying
the Laguerre ensemble.
 Although they have used the mathematical structure of this group to propose a
model to describe disordered metals, we believe that the hard edge of the
ensemble generated from the group of $\Omega$-matrices is actually more
appropriate to describe ballistic cavities.
\par
Ballistic cavities are of considerable current interest mainly because of
important recent breakthroughs in nanolithograpy, which has enabled its
construction in novel high-mobility semiconductor heterostructures \cite{1c10}.
These systems have elastic and inelastic mean free paths exceeding the device
dimensions at
 sufficiently low temperatures. As a consequence, transport in these structures
is dominated by scattering at the boundaries of the sample rather than the more
usual impurity scattering of mesoscopic metals. Two striking features of the
physics of these systems are weak-localization \cite{1c9} and universal
conductance fluctuations \cite{1c13,1c14}. Such typically mesoscopic phenomena
appear in ballistic cavities due to the chaotic nature of the boundary
scattering potential.
\par
In this work we shall mostly be concerned with a set-up that consists of an
interaction region of finite volume (the resonant cavity in a microwave
experiment or the ballistic microstructure in mesoscopic physics) connected to
two reservoirs by free propagation regions (wave guides for microwaves or
perfectly conducting leads for electron waves) in which asymptotic scattering
channels can be defined. The interaction region is assumed to  "trap" the
incoming waves by irregular boundary scattering thereby driving the system to a
regime where the ray optic limit
 (or classical dynamics) is dominated by classical chaos. We shall call such a
set-up an open ballistic cavity.
In particular, we study the problem of parametric correlations in open
ballistic cavities, that is, the response of
 the random spectra to an external adiabatic perturbation. Our main objective
is twofold: first, we want to extend the PRMT to describe open quantum chaotic
systems, such as ballistic cavities coupled to external reservoirs; and second
we want to understand how the theory changes when the region of physical
interest is at the hard edge of the support of the spectrum. The physics of the
soft edge has been discussed elsewhere \cite{amsm2}.
\par
Parametric correlations in open mesoscopic systems has recently been the
subject of
many works \cite{fhram,Rau,amsm3}. A concise account of some of our results has
been
presented previously \cite{amsm3}. In this work we provide much more details of
the calculations, a different interpretation of the hydrodynamic limit and new
results.

In section II we discuss the S-matrix approach to ballistic systems. In section
III, we review some of the properties of the group of $\Omega$-matrices of
Slevin and Nagao. In addition, we derive a diffusion equation generated by a
random walk on the $\Omega$-matrix manifold. We propose a modified version of
this equation as a Brownian motion model for parametric correlations in open
ballistic cavities. In section IV, the Brownian motion
 model is used to derive exact non-perturbative expressions for the
parameter-dependent two-point correlation function at the hard edge of the
spectrum. We demonstrate that after appropriate rescaling this function becomes
universal. In section V we show that in the hydrodynamic limit our theory
predicts a Lorentzian-squared behaviour for the correlator of conductance
fluctuations of ballistic cavities in agreement with semiclassical
calculations. Finally, we derive a formula for the power spectrum of an
arbitrary linear statistic on the transmission eigenvalues. A summary and
conclusions are presented in section VI.
\section{S-matrix and Ballistic Systems}
\par
It is now well established that the most convenient description of quantum
transport in open ballistic cavities is given by the S-matrix \cite{2c6,2c15}.
 By definition the $S$-matrix relates the incoming flux amplitudes $I_l$ and
$I_r$ to the outgoing ones $O_l$ and $O_r$ through the formula
\begin{equation}
S\pmatrix{I_l \cr I_r \cr}=\pmatrix{O_l \cr O_r \cr},
\label{smat1}
\end{equation}
where the subscripts $l$ and $r$ denote the left and the right sides of the
sample respectively. Current conservation implies that $S$ is unitary:
$S^{-1}=S^\dagger$. A general and rather simple explicit parametrization for
$S$ is \cite{2c6}
\begin{equation}
S=\pmatrix{r & t' \cr
           t & r' \cr}=\pmatrix{u^{(1)}  &  0   \cr
           0  &  u^{(2)} \cr}
\pmatrix{-\sqrt{1-\tau}  &  \sqrt{\tau}   \cr
         \sqrt{\tau}    &  \sqrt{1-\tau} \cr}
\pmatrix{v^{(1)}  &  0   \cr
         0  &  v^{(2)} \cr},
\label{smat2}
\end{equation}
where $r$ and $r'$ are $N \times N$ reflection matrices; $t$ and $t'$ are $N
\times N$ transmission matrices; $u^{(i)} (i=1,\dots,4)$ are $N \times N$
unitary matrices and $\tau$ denotes an $N \times N$ diagonal matrix with real
eigenvalues $0 \le \tau_\alpha \le 1 (\alpha=1,2,\dots,N)$, which are called
transmission eigenvalues. The total transmission coefficient can be written as
\begin{equation}
T\equiv{\rm tr} t t^{\dagger}=\sum_{\alpha=1}^N \tau_\alpha.
\label{trans}
\end{equation}
Therefore, the Landauer-B\"uttiker conductance is simply
\begin{equation}
G=G_0\sum_{\alpha=1}^N \tau_\alpha.
\label{cond2}
\end{equation}
The remarkable simplicity of this formula, in comparison with the
Kubo-Greenwood expression \cite{2c7} for the conductance obtained directly from
linear response theory, highlights the advantage of the Landauer-B\"uttiker
formalism. This, however, cannot be the end of the story because in principle
the transmission eigenvalues,
 $\tau_\alpha$, depend in a very complicated way on the underlying quantum
dynamics induced by the scattering
 mechanism. Fortunately, in some cases of interest such as ballistic cavities
and quasi-one-dimensional disordered conductors the complexity of the dynamics
turns out to be such that most details of microscopic origin are irrelevant and
the joint probability distribution of the $\tau_\alpha$'s can be determined
from symmetry arguments which are well known in the context of random matrix
theory \cite{2c8}.

\par
In the case of ballistic cavities,
it has been shown \cite{2c9} that the $S$-matrix fluctuates as a function of
the incident momentum due to multiple overlapping resonances in the cavity.
There is considerable numerical evidence \cite{2c10,2c11,2c12,2c13} to support
the assumption that through this fluctuations the $S$-matrix covers its
manifold with uniform probability.
 With this ergodic hypothesis in mind, we can study the fluctuations in the
$S$-matrix by means of an ensemble of random matrices with a uniform
distribution.
 So, $dP_\beta(S)$, which is the probability to find $S$ in the volume element
$dS$, is proportional to the Haar measure \cite{2c14} of the group of
S-matrices. In terms of the parametrization (\ref{smat2}) one finds, after
integration over the eigenvectors distributions \cite{2c6,2c15}
\begin{equation}
dP_\beta(\tau)\equiv d\mu_\beta(\tau)=C_{N,\beta} J_\beta(\tau)\prod_c
\tau_c^{(\beta-2)/2} \prod_a d\tau_a ,
\label{dP}
\end{equation}
where $C_{N,\beta}$ is a normalization constant and $\beta$ is a symmetry
index, whose value is $\beta=1$ for systems with time-reversal symmetry
(T-symmetry) in the absence of spin-orbit scattering, $\beta=2$ for systems
without T-symmetry and $\beta=4$ for systems with T-symmetry in the presence of
spin-orbit scattering. The factor $J_\beta(\tau)\equiv \prod_{a<b}|\tau_a
-\tau_b|^\beta$ is ultimately responsable for transmission eigenvalue repulsion
in the ensemble.
\par
At this stage it is convenient to make the following change of variables,
$\tau_i=1/(1+\nu_i^2)$, with $0 \le \nu_i < \infty$. From Eq. (\ref{dP}) we
find the following joint probability density for the variables $\{\nu_i\}$
\begin{equation}
P(\nu)=Z^{-1} \exp(-\beta {\cal H}),
\label{2.P1}
\end{equation}
where
\begin{equation}
{\cal H}=-{1 \over 2} \sum_{i \ne j} Q(\nu_i,\nu_j) +
\sum_i V(\nu_i),
\label{2.P2}
\end{equation}
\be
Q(\nu,\nu')=\ln |\nu^2-\nu'^2|,
\label{2.P3}
\ee
and
\be
V(\nu)=\left(N+{2-\beta \over 2\beta}\right)\ln (1+\nu^2)-{1 \over \beta}
\ln \nu.
\label{2.P4}
\ee
Note that $P(\nu)$ has the form of a Gibbs distribution and ${\cal H}$ plays
the role of a Hamiltonian of classical particles with logarithmic pairwise
repulsion, $Q(\nu,\nu')$, and a confining potential, $V(\nu)$.
The dimensionless Landauer-B\"uttiker conductance in this new set of variables
reads
\begin{equation}
g=\sum_{i=1}^N g(\nu_i),
\label{glin}
\end{equation}
where $g(\nu)=(1+\nu^2)^{-1}$. In RMT observables of this form are called
linear statistics \cite{2c8} because products of different eigenvalues do not
appear in their defining expressions. Consequently, their statistical
properties are the simplest. In particular, the average and variance of Eq.
(\ref{glin}) are simply
\begin{equation}
\bigl\langle g\bigr\rangle=\int_0^\infty g(\nu)\rho(\nu) d\nu,
\label{gave}
\end{equation}
\begin{equation}
{\rm var}(g)=\int_0^\infty g(\nu)g(\nu')S(\nu,\nu')d\nu d\nu',
\label{gvar}
\end{equation}
in which  $\rho(\nu)$ and $S(\nu,\nu')$ are the average level density and
two-point correlation function respectively. There is a very powerful way of
calculating $S(\nu,\nu')$ which has been developed by Beenakker \cite{2c16}. It
is based on the following identity
\begin{equation}
S(\nu,\nu')=-{1 \over \beta} {\partial \rho(\nu) \over \partial V(\nu')
},
\label{K}
\end{equation}
which can be easily verified. For $N \gg 1$, one can show that
\begin{equation}
V(\nu)=\int_0^\infty \rho(\nu')Q(\nu,\nu')  d\nu'.
\label{V}
\end{equation}
Using (\ref{V}) the functional derivatives in (\ref{K}) can be performed and we
find
\begin{equation}
S(\nu,\nu')=-{1 \over \beta}Q^{-1}(\nu,\nu')=
{1 \over \pi^2 \beta}{\partial^2 \over \partial \nu
\partial \nu'} \ln \left[{\nu+\nu' \over
\nu-\nu' } \right]+O(1/N).
\label{K0}
\end{equation}
A remarkable consequence of this technique is the conclusion that the leading
expression for $S(\nu,\nu')$, in an expansion in inverse powers of $N$, is
independent of the potential $V(\nu)$. Therefore, the variance of a linear
 statistic, such as the conductance (see Eq. (\ref{gvar})), is the same for all
random matrix ensembles with an edge at the origin of the spectrum (since the
$\nu$'s are all non-negative) and with an eigenvalue repulsion
 potential equal to $Q(\nu,\nu')$. Consequently, if we are only interested in
studying the dominant contribution to fluctuations in mesoscopic observables,
we are free to choose the functional dependence of $V(\nu)$ that we find
convenient. It turns out that the simplest choice for $V(\nu)$ can be obtained
from (\ref{2.P4}) by means of the rescaling, $\nu \to \nu/N$, and by taking the
large $N$ limit keeping $\nu$ fixed. We then get
\be
V(\nu)={1 \over N} \nu^2- {1 \over \beta} \ln \nu.
\label{2.P5}
\ee
We remark that Eqs. (\ref{2.P1}), (\ref{2.P2}) and (\ref{2.P3}) with $V(\nu)$
given by (\ref{2.P5}) constitute the Laguerre ensemble of random matrices
\cite{2c17}. This ensemble has a number of interesting universal properties
which will be discussed in section III, where it is used to build a model of
parametric correlations in ballistic cavities.
\par
We conclude this section by giving the leading terms in the expansions of
$\langle g \rangle$ and ${\rm var}(g)$ in inverse powers of $N$
\begin{equation}
\langle g \rangle ={1 \over 2} N + {\beta-2 \over 4\beta}+O(1/N),
\label{g}
\end{equation}
\begin{equation}
{\rm var}(g)={1 \over 8\beta}+O(1/N).
\label{g2}
\end{equation}
The leading term in Eq. (\ref{g}) is a direct consequence of the ergodic
hypothesis, since it implies that the average transmission and reflection
probabilities are equal. The second term in Eq. (\ref{g}) corresponds to the
 weak-localization correction, which as is due to constructive interference of
time-reversed backscattering trajectories. Eq. (\ref{g2}) is an illustration of
the remarkable phenomenon of universal conductance fluctuations, which is a
clear signature of the non-self-averaging nature of observables in quantum
transport theory.

\section{The Group of $\Omega$-Matrices}
In this section we introduce the group of $\Omega$-matrices as an abstract
mathematical structure, which, as we shall see, is very convenient for studying
the Laguerre ensemble. We remark that our approach differs considerably
 in philosophy from that used by Slevin and Nagao \cite{5c31}. They have
motivated the group of $\Omega$-matrices by relating it to the group of
transfer matrices of a disordered conductor. We believe that to make such a
connection
 from the outset is unnecessary and might lead to misinterpretations. The group
of $\Omega$-matrices is interesting in its own right as a powerful way of
describing statistical properties of the Laguerre ensemble, which we regard
 as a tractable mathematical model for studying local eigenvalue correlations
in systems with a hard edge in the spectrum. Furthermore, as we discussed in
section II, the formula for the variance of linear statistics derived from the
Laguerre ensemble applies quite generally to any maximum-entropy ensemble with
a hard edge, most notably, the ensemble describing ballistic cavities.
\par
By definition, a general complex matrix $\Omega$  satisfies
\be
\Sigma_z \Omega \Sigma_z=-\Omega.
\label{5.rel1}
\ee
If the system has T-symmetry we have the additional relation
\be
\Omega^*=\Sigma_x \Omega \Sigma_x.
\label{5.rel2}
\ee
The matrices $\Sigma_z$ and $\Sigma_x$ are given by
\be
\Sigma_z=\pmatrix{1  &  0   \cr
         0  &  -1 \cr},
\ee
and
\be
\Sigma_x=\pmatrix{0  &  1   \cr
         1  &  0 \cr},
\ee
where $1$ is the $N \times N$ unit matrix for orthogonal and unitary systems,
and it is the $N \times N$ unit quaternion matrix for symplectic ensembles.

One can show that a general matrix $\Omega$ satisfying (\ref{5.rel1}) and
(\ref{5.rel2}) can be written in the form
\begin{equation}
\Omega=\pmatrix{u  &  0   \cr
           0  &  v \cr}
\pmatrix{0  &  \nu   \cr
         \nu    &  0 \cr}
\pmatrix{u^\dagger&  0   \cr
         0  &  v^\dagger \cr},
\label{5.omega}
\end{equation}
where $u$ and $v$ are  $N \times N$ unitary matrices for unitary and orthogonal
systems and they are $N \times N$ quaternion unitary matrices for symplectic
systems. The matrix $\nu$ is a diagonal matrix with real eigenvalues. One can
easily check that the matrices $\Omega$ form a group under addition, which we
call ${\cal G}^{\beta}(\Omega,+)$.
\par
Let $\Omega_t$ denote a point on the $\Omega$-matrix manifold that is generated
from a random walk defined by
\be
\Omega_t=\sum_{s=1}^t \Omega_s(\delta t),
\label{5.walk}
\ee
where $\delta t$ is the step of the walk and $\Omega_s(\delta t)$ is a random
matrix belonging to ${\cal G}^{\beta}(\Omega,+)$. The simplest choice for
$\Omega_s(\delta t)$ is
\be
\Omega_s(\delta t)=2i\sqrt{\delta t}\pmatrix{0  &  y   \cr
           -y^\dagger  &  0 \cr},
\label{5.step}
\ee
where $y$ is a complex random matrix with vanishing average and with second
moment given by
\begin{equation}
\langle y^{}_{ab}y^*_{cd} \rangle={ 1 \over \gamma
(N+1)}(\delta_{ac}\delta_{bd} +\delta_{ad}\delta_{bc}),
{}~~~~~~~~~~\beta=1,
\label{5.y2.1}
\end{equation}
\begin{equation}
\langle y^{}_{ab}y^*_{cd} \rangle={ 1 \over \gamma
N}\delta_{ac}\delta_{bd},~~~~~~~~~~\beta=2,
\label{5.y2.2}
\end{equation}
\begin{equation}
\langle y^{}_{ab\alpha}y^*_{cd\alpha'} \rangle={ 1 \over 2\gamma
(2N-1)}(\delta_{ac}\delta_{bd}
+(2\delta_{\alpha,0}-1)\delta_{ad}\delta_{bc})\delta_{\alpha,\alpha'},
{}~~~~~~~~~~\beta=4,
\label{5.y2.3}
\end{equation}
in which $\gamma$ is a constant. Note that for systems with symplectic symmetry
($\beta=4$) we have used the fact that $y$ is a quaternion matrix and thus any
of its matrix elements can be written as $y_{ab}=\sum_{\alpha=0}^3
y_{ab\alpha}e_\alpha$, where $e_\alpha$ are the quaternion units.
\par
We can rewrite $(\ref{5.walk})$ as $\Omega_{t+\delta
t}=\Omega_t+\Omega_s(\delta t)$, which in view of $(\ref{5.omega})$ implies, to
order $O(\delta t)$, the stochastic process
\be
\delta \nu_i=i \sqrt{\delta t} (A_{ii}-A_{ii}^\dagger)/\nu_i +
{2\delta t \over  \gamma (\beta N+2-\beta)}\left[{1 \over \nu_i} +
2\beta\sum_{j (\ne i)}
{\nu_i \over \nu_i^2-\nu_j^2}
\right],
\label{5.nu}
\ee
where $A_{ij}\equiv(u^\dagger y v \nu)_{ij}$ and $\nu_i\equiv (\nu)_{ii}$. If
we take the continuum limit of (\ref{5.nu}) we find the following Fokker-Planck
equation for the probability distribution $W(\nu,\tau)$
\begin{equation}
\gamma{\partial W \over \partial \tau}=
\sum_i \left( -{\partial \over \partial \nu_i}D_i +
{\partial^2 \over \partial \nu_i^2} \right) W,
\label{5.W}
\end{equation}
where $\tau=2t/(\beta N+2-\beta)$ and
\be
D_i={1 \over \nu_i} + 2\beta \sum_{j (\ne i)}
{\nu_i \over \nu_i^2-\nu_j^2}.
\label{5.drift1}
\ee
It is interesting to note that the transformation $W=JP$ implies that $P$
satisfies
\begin{equation}
\gamma {\partial P \over \partial \tau}=
{1 \over J}\sum_i  {\partial \over \partial \nu_i}J
{\partial \over \partial \nu_i } P = \nabla^2 P,
\label{5.P}
\end{equation}
where $J=\prod_i \nu_i \prod_{i<j}|\nu_i^2-\nu_j^2|^\beta$ and $\nabla^2$ is
the Laplace-Beltrami operator of the $\Omega$-matrix manifold.
\par
In the analysis of Slevin and Nagao \cite{5c31}, the distribution of
$\Omega$-matrices was determined by a maximum-entropy criterion to be the
Laguerre ensemble
\be
W^L(\nu)=C_{N,\beta} \prod_{i<j} \left|\nu_i^2-\nu_j^2\right|^\beta
\prod_{i=1}^N \nu_i^{\beta \alpha+1} \exp(-{\beta \over 2} c \nu_i^2),
\label{5.lag}
\ee
where $C_{N,\beta}$ and $c$ are constants, $\alpha=0$ and $\beta$ is the usual
symmetry index. This result suggests, in analogy with transport theory, that
our Brownian motion in the $\Omega$ manifold must satisfy the requirement that
when $\tau \to \infty$, $W(\nu,\tau) \to W^L(\nu)$, that is the distribution
function evolves to a configuration that maximizes the entropy. This can be
achieved simply by adding to $D_i$ a term corresponding to a confining force so
that (\ref{5.W}) becomes
\be
\gamma{\partial W\over \partial \tau} =
\sum_{i=1}^{N} {\partial \over \partial \nu_i}
\left(W{\partial \Phi \over \partial \nu_i}+{1\over \beta}{\partial W \over
\partial \nu_i}\right),
\label{5.6a}
\ee
in which
\be
\Phi(\nu)=-(\alpha+{1 \over \beta})\sum_i \ln \nu_i + {c \over 2} \sum_i
\nu_i^2
-\sum_{i<j} \ln\left|\nu_i^2-\nu_j^2\right|.
\label{5.6b}
\ee
where $\beta=1$ and $\alpha=0$. Eq. (\ref{5.6a}) is the main result of this
section. It can be regarded as representing a Brownian motion of $N$ classical
particles at positions $\{\nu_i(\tau)\}$ moving in a viscous fluid with
friction coefficient $\gamma^{-1}$ at temperature $\beta^{-1}$. The parameter
$\tau$ is a fictitious time
 which we shall relate to the perturbation parameter, so that (\ref{5.6a})
becomes a model to describe parametric correlations in the Laguerre ensemble.
In the framework of PRMT it has been explicitly demonstrated that all the
universal functions can be obtained from a Brownian motion model if one assumes
that $\tau=(\delta u^2)$ \cite{5c17,5c18,5c19,5c20}, where $\delta u$ is a
parameter characterizing the strength of the external perturbation. Motivated
by this result, we assume the same relation to apply to our model. We stress
that the consistency of this assumption has recently been confirmed by
 explicit calculations of a similar model using supersymmetry \cite{5c36}.
 The apparently artificial parameter $\alpha$ has been introduced to allow the
classification and understanding of the universal expressions, which we derive
in section IV, in the more general context of Laguerre ensembles. Note that Eq.
(\ref{5.lag}) with $\alpha >-1$ characterizes the most general Laguerre
ensemble of random matrices.

\section{Universal Correlations at the Hard Edge}
A crucial difference between open ballistic systems and metallic conductors is
that, in the former, the joint probability distribution of the transmission
eigenvalues can be directly obtained from a maximum-entropy principle,
 whilst as shown in Refs. \ \onlinecite{4c16}, \ \onlinecite{amsm4} and \
\onlinecite{amsm5} this is not the case for disordered conductors. A direct
consequence of this fact is that the fluctuations in transport observables of
 ballistic systems are completely characterized by the universal logarithmic
repulsion of conventional random matrix ensembles, like the Laguerre one. An
outstanding common feature of the ensembles of random matrices
 appropriate to describe open ballistic systems and disordered conductors,
though, is that the spectrum of transmission eigenvalues has a hard edge, since
the eigenvalues are all non-negative. This is, as we discussed in the
Introduction, also a feature of the Laguerre ensemble. In this section we shall
demonstrate that the presence
 of this hard edge implies the existence of universal parametric correlations
of a new kind.
\subsection{Exact Mapping onto a Schr\"odinger Equation}
\par
Our starting point is Eq. (\ref{5.6a})-(\ref{5.6b}), which we rewrite here in
the form
\be
\gamma{\partial W\over \partial \tau} = {\cal L}_{FP} W,
\label{5.wfp}
\ee
where ${\cal L}_{FP}$ is the Fokker-Planck operator, given by
\be
{\cal L}_{FP}=\sum_i {\partial \over \partial \lambda_i}
\left(
-D^{(1)}_i+{\partial \over \partial \lambda_i}D^{(2)}_i
\right),
\label{5.lfp}
\ee
where $\lambda_i\equiv \nu_i^2$ and
\be
D^{(1)}_i=4\sum_{j \ne i} {\lambda_i \over \lambda_i-\lambda_j}
-2c\lambda_i+2\alpha+4/\beta
\label{5.drift}
\ee
while
\be
D^{(2)}_i={4\lambda_i \over \beta} \delta_{ij}.
\label{5.diff}
\ee
At this stage, it is useful to perform the following transformation \cite{5c37}
\be
P(\lambda,\tau)=\exp\left(-{1 \over 2}\Phi(\lambda)\right)
\Psi(\lambda,\tau),
\label{5.trans}
\ee
which implies that $\Psi(\lambda,\tau)$ satisfies the evolution equation
\be
{\partial \Psi \over \partial \tau}={\cal L} \Psi,
\label{5.sch}
\ee
where ${\cal L}={\cal L}^\dagger$ is given by
\be
{\cal L}=e^{\Phi /2}{\cal L}_{FP}e^{-\Phi /2}.
\label{5.L}
\ee
{}From (\ref{5.6b}), (\ref{5.lfp})-(\ref{5.diff}) and (\ref{5.L}) we obtain
${\cal L}=B-{\cal H}$, where
\be
{\cal H}=\omega
 \sum_i \biggl[
-{\partial \over \partial r_i}r_i{\partial \over \partial r_i}+\left({\beta
\over 4}\right)^2(r_i+{\alpha^2 \over r_i})+
{\beta(\beta-2) \over 4}\sum_{j(\ne i)} {r_i \over (r_i-r_j)^2} \biggr ],
\label{5.H}
\ee
$B=N c(2+\beta(N+\alpha-1))/(2\gamma)$, $\omega=4c/(\beta \gamma)$ and
$r_i=c\lambda_i$. The operator ${\cal H}$ can be regarded as a one dimensional
Hamiltonian of fictitious hard-core particles, for which it is convenient to
assign fermionic statistics. We remark that it is a general property
of the mapping (\ref{5.trans}) that equilibrium fluctuations of the classical
Brownian particles correspond to quantum mechanical ground-state fluctuations
of these fictitious fermions. Note that for $\beta=2$ the two-body interaction
term vanishes and ${\cal H}$ becomes the Hamiltonian of a system of free
fermions. In this case we can rewrite ${\cal H}$ as
\be
{\cal H}=\sum_{p=0}^{\infty} \varepsilon_p c^{\dagger}_pc^{}_p,
\label{5.free}
\ee
where $\varepsilon_p=(2c/\gamma)(p+(\alpha+1)/2)$ and $c^{\dagger}_p$
$(c^{}_p)$ creates (annihilates) a fermion with quantum number $p$. It is
convenient at this stage to introduce the field operators $\psi(\lambda)$ and
$\psi^{\dagger}(\lambda)$, where $\psi(\lambda)\equiv \sum_p \tilde
\phi_p(\lambda)c_p^{}$ and $\psi(\lambda)\equiv \sum_p \tilde
\phi_p(\lambda)c_p^{\dagger}$, in which
\be
\tilde \phi_p(\lambda)=\left({c^{1+\alpha} p! \over
\Gamma(p+1+\alpha)}\right)^{1/2}
\lambda^{\alpha/2}e^{-c \lambda/2}L^\alpha_p(c \lambda),
\label{5.1part}
\ee
with $L^\alpha_p(x)$ denoting the associate Laguerre polynomial.
\par
With this notion, we turn to the calculation of some quantities of physical
interest, namely the average level density and the two-point correlation
function for level-density fluctuations. In second quantized language the local
density operator is defined as
\be
\hat n(\lambda,\tau)=
\psi^\dagger(\lambda,\tau) \psi(\lambda,\tau)
\label{5.n}
\ee
where
\be
\psi^\dagger(\lambda,\tau)=e^{{\cal H}\tau} \psi^\dagger(\lambda)e^{-{\cal
H}\tau}
\ee
and
\be
\psi(\lambda,\tau)=e^{{\cal H}\tau} \psi(\lambda)e^{-{\cal H}\tau}.
\ee
Let $|0\rangle$ denote the $N$-fermion ground-state of (\ref{5.free}). Then the
average level density is simply
\be
\tilde \rho(\lambda,\tau)=\langle 0| \hat n(\lambda,\tau) | 0 \rangle
=\sum_{p=0}^{N-1} \tilde \phi_p(\lambda)^2=
\tilde \rho(\lambda),
\ee
which is independent of $\tau$, as expected, since the Brownian particles are
in a stationary state.
\par
The two-point function for level-density fluctuations is defined as
\be
\tilde S(\lambda,\lambda',\tau)\equiv
\langle 0| \hat n(\lambda,\tau) \hat n(\lambda',0) | 0 \rangle
-\tilde \rho(\lambda) \tilde \rho (\lambda).
\ee
Using (\ref{5.n}) and Wick's theorem we find
\bea
\tilde S(\lambda,\lambda',\tau)=
&G_0(\lambda,\lambda',\tau)\sum_{q=0}^{N-1} \tilde\phi_q(\lambda)
\tilde\phi_q(\lambda')e^{\varepsilon_q \tau}-\nonumber \\
&\sum_{p=0}^{N-1} \sum_{q=0}^{N-1}\tilde\phi_p(\lambda)
\tilde\phi_p(\lambda')\tilde\phi_q(\lambda)
\tilde\phi_q(\lambda')e^{(\varepsilon_q-\varepsilon_p) \tau},
\label{5.S}
\eea
where
\be
G_0(\lambda,\lambda',\tau)={c \exp \left( -{c \over 2}(\lambda+\lambda')\coth
(\omega \tau /2) \right)\over 2 \sinh (\omega \tau /2)}
I_\alpha\left({c\sqrt{\lambda \lambda'} \over \sinh (\omega \tau /2)} \right)
\label{5.G}
\ee
and $I_\alpha(x)$ is a Bessel function. Equations (\ref{5.S}) and (\ref{5.G})
can be considered as a generalization to finite $\tau$ of the standard result
of RMT \cite{2c8}, which we recover by taking the limit $\tau \to 0$, giving
\be
\tilde S(\lambda,\lambda',0)= \delta(\lambda-\lambda')\tilde
\rho(\lambda)-\left(\tilde K(\lambda,\lambda')\right)^2,
\ee
where
\be
\tilde K(\lambda,\lambda')= \sum_{p=0}^{N-1} \tilde\phi_p(\lambda)
\tilde\phi_p(\lambda').
\ee
In the next section we show how to rescale the average level density and the
two-point correlation function at the hard edge of the spectrum.

\subsection{Rescaling Physical Quantities at the Hard Edge}
\par
In order to obtain the new universal behaviour anticipated in the Introduction
we need to rescale all the quantities of physical interest at the hard edge of
the spectrum, that is at $\lambda=0$. This is done more easily if we
reintroduce the original variables $\nu_i=\sqrt{\lambda_i}$. So, the average
level density becomes
\be
\rho(\nu)=\sum_{p=0}^{N-1}\phi_p(\nu)^2,
\label{5.dens}
\ee
where $\phi_p(\nu)\equiv \sqrt{2\nu}\tilde \phi(\nu^2)$. For large $N$ we can
use the asymptotic \cite{5c38}
\be
\phi_N(\nu) \simeq (2c\nu)^{1/2} J_\alpha\left(2(N c)^{1/2} \nu\right),
\label{5.asymp}
\ee
where $J_\alpha(x)$ is a Bessel function. Inserting (\ref{5.asymp}) into
(\ref{5.dens}) and taking $N \to \infty$ and
$c \to 0$ such that $2\sqrt{Nc} \to \rho_0 \pi$ we find
\be
\rho(\nu)=\pi^2\rho_0^2\int_0^1 ds s \nu J_\alpha^2(\pi \rho_0 \nu s),
\ee
which is exactly the result obtained in Refs. \cite{5c27,5c28,5c29} for the
average level density rescaled at the hard edge of the spectrum. Note that the
constant $\rho_0$ is, as can be seen from the relation $\rho(\nu)\simeq
\rho_0$, valid for large $\nu$, just the bulk average level density.
\par
A similar calculation yields for the two-point function
\bea
S(\nu,\nu',\delta u)=&\pi^4\rho_0^4 \int_0^1 ds\int_1^\infty ds' ss'\nu\nu'
J_\alpha(\pi \rho_0 \nu s)J_\alpha(\pi \rho_0 \nu' s)J_\alpha(\pi \rho_0 \nu
s') \nonumber \\
&J_\alpha(\pi \rho_0 \nu' s')\exp\left({\pi^2\rho_0^2 \over 2 \gamma}\delta
u^2(s^2-
{s'}^2)\right).
\label{5.2pt}
\eea
This is the central result of this section. Note that, for $\delta u=0$, Eq.
(\ref{5.2pt}) reproduces known results \cite{5c27,5c28,5c29} for the two-point
correlator of ensembles with a hard edge.
Therefore, we believe that Eq. (\ref{5.2pt}) represents an extension of these
RMT results to account for the dispersion of the eigenvalues as a function of
the perturbation parameter. Bearing in mind physical applications, consider now
the particular case of the ensemble describing open ballistic sytems, which can
be mapped (see section 2.2) onto a Laguerre ensemble with $\alpha=0$.
Following Ref. \ \onlinecite{5c15}  we make the rescalings $\hat \nu=\rho_0
\nu$, $\hat \nu'=\rho_0 \nu'$, $\delta \hat u^2=\delta u^2 \rho_0^2/\gamma$ and
$\hat S(\hat \nu,\hat \nu',\delta \hat u)=S(\nu,\nu',\delta u)/\rho_0^2$. We
can see that $\hat S(\hat \nu,\hat \nu',\delta \hat u)$ is a function that is
independent of any physical
 parameter, such as the Fermi velocity or size of the sample, and therefore can
be regarded as a universal
 characterization of quantum chaotic scattering in such systems.  For the more
general case of Laguerre ensembles, with $\alpha > -1$, we can see that the
parameter $\alpha$ labels new universality classes characteristic of systems
with a hard edge in the
spectrum. Finally, we remark that Eq. (\ref{5.2pt}) has recently been derived,
for $\alpha=0$, by Andreev, Simons and Taniguchi \cite{5c36} using the
supersymmetry technique. In addition, they have demonstrated that the universal
behaviour at the hard edge extends to another kind of two-point function.

\par
Another quantity of some interest is the distribution of level velocities
defined as
\be
K(v)\equiv \langle \delta(v-d\nu/du) \rangle.
\ee
One can show that \cite{5c39}
\be
K(v)=\lim_{\delta u \to 0} {\delta u \over \rho(\nu)} S(\nu,\nu+v \delta
u,\delta u).
\ee
Now using (\ref{5.S}) with $\tau=\delta u^2$ we find
\be
 S(\nu,\nu+v \delta u,\delta u)\simeq{\rho(\nu) \over \delta u} \left({\gamma
\over 2\pi}\right)^{1/2}\exp\left(-{\gamma v^2 \over
2}\right)
\ee
and therefore the distribution of level velocities is gaussian
\be
K(v)=\left({\gamma \over 2\pi}\right)^{1/2}\exp\left(-{\gamma v^2 \over
2}\right).
\label{5.K}
\ee
Remarkably, this is exactly the same result that one finds at the bulk of the
spectrum. We stress that this result, although simple, is by no means trivial,
since the hard edge rescaling procedure introduces considerable changes in
most physical quantities, such as the average level density and two-point
correlation function.
One interesting application of (\ref{5.K}) is associated with the
interpretation of the rescalings discussed right after Eq. (\ref{5.2pt}). The
quantity $C_0=\rho_0^2/\gamma$ can now, in view of (\ref{5.K}), be written as
\be
C_0=\rho_0^2 \langle v^2 \rangle,
\ee
so that the rescalings $\hat \nu=\rho_0 \nu$ and $\delta \hat u^2=\delta u^2
C_0$ mean that we get universal functions if we measure the levels in units of
the average bulk level spacing $\rho_0^{-1}$ and if the perturbation parameter
is rescaled by the mean square gradients of the levels $\sqrt{C_0}$.

\section{The Power Spectrum Formula}
A complete characterization of the fluctuations in transport observables in
mesoscopic systems consists of a description of both their magnitude and power
spectrum. One can get this information from the correlator
\be
F(\delta u)=\langle \delta A(\delta u) \delta A(0) \rangle,
\label{5.F}
\ee
where $A$ is an arbitrary transport observable and $\delta u$ parametrizes the
external perturbation on the system which drives the fluctuations. The power
spectrum $C(\omega)$ is defined as
\be
C(\omega)=\int_{-\infty}^\infty dt e^{i \omega t} F(t).
\label{5.C}
\ee
Microscopic diagrammatic calculations \cite{1c13,1c14} for disordered metals
suggest that $F(\delta u)$  decays as a power law as a function of $\delta u /
{\cal E}_c$, where ${\cal E}_c$ is a correlation parameter which sets the
typical scale of the spacings between the peaks and valleys in the values of
$A$, for a given realization, as a
 function of $\delta u$. This slow power law decay of $F(\delta u)$ indicates
that the system sustains some sort of long-range memory in the fluctuations of
the observable. This kind of behavior must be contrasted with the exponential
decay of Poisson processes and can be regarded as a manifestation of spectral
rigidity, a well known phenomenon in the theory of random matrices.
\par
It is well known \cite{amsm4,amsm5} that universal fluctuations in transport
observables of mesoscopic disordered conductors can be obtained  by simply
taking the hydrodynamic limit of the DMPK equation. We expect a similar
 procedure to apply to the power spectrum formula of open ballistic structures.
In section II, we have demonstrated
 that the Laguerre ensemble can be used to describe fluctuations in transport
observables of ballistic cavities.
 With this in mind we shall now discuss the hydrodynamic limit of Eq.
(\ref{5.6a}).
\par
Multiplying (\ref{5.6a}) by $\sum_i \delta(\nu-\nu_i)$, integrating over all
$\nu_i$'s and using the definition
\be
\rho(\nu,\tau)\equiv\left\langle \sum_i \delta(\nu-\nu_i)\right\rangle_\tau,
\ee
where $\langle\dots\rangle_\tau$ denotes an average over the distribution
$W(\nu,\tau)$, we get in the large $N$ limit the following evolution equation
for the average level density
\be
\gamma {\partial \rho(\nu,\tau) \over \partial \tau} \simeq
-{\partial \over \partial \nu}\left[ \rho(\nu,\tau) {\partial \over \partial
\nu}
\left( \int_0^\infty d\mu\rho(\mu,\tau)\ln\left|\nu^2- \mu^2\right|-{c \over
2}\nu^2\right)\right].
\label{5.heq}
\ee
One can see from (\ref{5.heq}) that the equilibrium density satisfies
\be
\int_0^\infty \rho_{eq}(\mu) \ln\left|\nu^2- \mu^2\right|
d\mu={c \over 2} \nu^2+{\rm const.}
\label{5.dseq}
\ee
The constant on the right hand side of (\ref{5.dseq}) is determined by the
normalization condition
\be
\int_0^\infty \rho_{eq}(\nu) d \nu=N
\label{5.cond}
\ee
One can easily verify that
\be
\rho_{eq}(\nu)=\cases{ (c/ \pi)\sqrt{{4N\over c}-\nu^2}
& for $0\le \nu\le 2(N/c)^{1/2}$ \cr
0 & for $\nu > 2(N/c)^{1/2}$ \cr},
\label{5.leveldensity}
\ee
is the solution of (\ref{5.dseq}) satisfying (\ref{5.cond}). Note that for
large $N$ at fixed $\nu$ one finds
\be
\rho_{eq}(\nu)\simeq {2 \over \pi} \sqrt{Nc}=\rho_0,
\ee
as expected. In the regime of universal mesoscopic fluctuations that we are
concerned with, one can safely consider the average level density, $\rho_0$, to
be much larger than the fluctuating part $\delta \rho(\nu,\tau)$. Therefore, it
makes sense to try and linearize Eq. (\ref{5.heq}) by writing $\rho(\nu,\tau)$
as
\be
\rho(\nu,\tau)=\rho_0+\delta\rho(\nu,\tau).
\label{5.lin}
\ee
Inserting (\ref{5.lin}) into (\ref{5.heq}) yields
\be
\gamma {\partial \delta \rho(\nu,\tau) \over \partial \tau} \simeq
-\rho_0{\partial^2 \over \partial \nu^2}
 \int_0^\infty d\mu\delta\rho(\mu,\tau)\ln\left|\nu^2- \mu^2\right|.
\label{5.hlin}
\ee
Since
\be
\ln\left|\nu^2- \mu^2\right|=-2\int_0^\infty {dk \over k} \cos k\nu \cos k \mu,
\ee
Eq. (\ref{5.hlin}) can be solved by means of Fourier cosine transform. We find
\be
\delta \rho(\nu,\tau)={2 \over \pi} \int_0^\infty dk
\delta \tilde \rho(k,0) \exp (-\pi k \rho_0 \tau/\gamma) \cos k \nu
\label{5.delrho}
\ee
The two-point function $S(\nu,\nu',\delta u)$ can be obtained from
(\ref{5.delrho}) through the identity
\be
S(\nu,\nu',\delta u)=\left \langle \delta \rho(\nu,\tau)\delta \rho(\nu',0)
\right \rangle_{eq}
\label{5.Sdef}
\ee
where $\langle\dots\rangle_{eq}$ stands for an average over the equilibrium
distribution (\ref{5.lag}).
\par
It is useful to define the following double Fourier cosine transform
\be
\tilde S(k,\delta u) = \int_0^\infty d\nu \int_0^\infty d\nu'
S(\nu,\nu',\delta u) \cos k\nu \cos k \nu'
\label{5.Sk}
\ee
so that we get from (\ref{5.Sdef})
\be
\tilde S(k,\delta u) = \left \langle \delta \tilde \rho(k,\delta u^2)\delta
\tilde \rho(k,0)
\right \rangle_{eq}=\exp(-\pi k \rho_0 \delta u^2/\gamma) \tilde S(k,0)
\ee
The Fourier component $\tilde S(k,0)$ can be obtained directly from Eq.
(\ref{K0}), which together with  (\ref{5.Sk}) yields
$\tilde S(k,0)=k/2\beta$, so that
\be
\tilde S(k,\delta u)={k \over 2 \beta} \exp(-\pi k \rho_0 \delta u^2/\gamma).
\label{5.Sku}
\ee
Let $A=\sum_i a(\nu_i)$ denote an arbitrary observable, which can be expressed
as a linear statistic. Then from (\ref{5.F}) and (\ref{5.Sdef}) we get for the
correlation function
\be
F(\delta u)=\int_0^\infty d\nu \int_0^\infty d\nu' a(\nu) a(\nu')
S(\nu,\nu',\delta u) ,
\ee
which by virtue of (\ref{5.Sku}) yields the formula
\be
F(\delta u) ={2 \over \beta \pi^2}\int_0^\infty dk k
\exp(-\pi k \rho_0 \delta u^2/\gamma) \tilde a^2(k),
\label{5.forF}
\ee
where
\be
\tilde a(k)=\int_0^\infty \cos k \nu a(\nu) d\nu.
\ee
Finally, from (\ref{5.forF}) we can obtain the power spectrum formula
\be
C(\omega)={2 \over \beta \pi^2}\left({\gamma \over \rho_0}\right)^{1/2}
\int_0^\infty{\tilde a^2(k) \over \sqrt{k}} \exp\left(-
{\gamma \omega^2 \over 4 \pi k \rho_0}\right) dk.
\label{5.forC}
\ee
Equations (\ref{5.forF}) and (\ref{5.forC}) are the principal result of this
section.
\par
As an application, which also serves as a test of our results, we consider the
two-probe Landauer-B\"uttiker dimensionless conductance (Eq. (\ref{glin}))
\be
g=\sum_i{1 \over 1+\nu^2_i}.
\ee
We find
\be
\tilde a_g(k)={\pi \over 2} e^{-k},
\ee
thus the correlator of Eq. (\ref{5.forF}) gives
\be
F(\delta u)={1 \over 8\beta }\left( 1+ \delta u^2/{\cal E}^2_c
\right)^{-2},
\label{5.gF}
\ee
where ${\cal E}_c=\sqrt{2\gamma/\pi \rho_0}$ is the correlation parameter.
For the power spectrum we get
\be
C(\omega)={\pi{\cal E}_c \over 16 \beta} e^{-\omega{\cal E}_c}
(1+\omega{\cal E}_c).
\label{5.gC}
\ee
We would like to stress that Eqs. (\ref{5.gF}) and (\ref{5.gC}), for $\beta=2$,
are in complete agreement with independent calculations based on semiclassical
quantization \cite{1c16}.
\section{Summary and Conclusions}
In this work we have studied the effects of an  external adiabatic perturbation
on the transmission eigenvalue correlations of open ballistic cavities.
\par
In particular, we have derived a Brownian motion model to describe dynamic
fluctuations in the Laguerre ensemble of random matrices, which we have
proposed as a model for parametric correlations of transmission eigenvalues in
ballistic cavities . This model has enabled us to obtain explicit
non-perturbative expressions for the two-point
 function of level density fluctuations at the hard edge of the spectrum. We
have shown that after appropriate
 rescaling this function becomes system independent and can therefore be used
as a signature of quantum chaos in open ballistic cavities. In the hydrodynamic
limit, we have demonstrated that the two-point parametric correlator
 of mesoscopic conductance fluctuations in ballistic cavities is a
Lorentzian-squared, in agreement with semiclassical calculations. We have also
obtained a formula for the power spectrum of the fluctuations of an arbitrary
linear statistic in such systems.

Parametric correlations in the S-matrix ensemble has recently been discussed in
Refs. \ \onlinecite{fhram} and  \ \onlinecite{Rau}. They obtained exponential
decay for the
conductance correlator, which disagrees with equation (\ref{5.gF}) and
consequently with microscopic semiclassical
 calculations. It is not understood at
the moment why the S-matrix approach, which was so successful in describing
non-parametric fluctuations, should have a hydrodynamical limit for parametric
correlations that disagrees with semiclassical results. We remark that
 our $\Omega$-matrix approach, on the other hand, although phenomenological in
nature, has the advantage of containing the correct semiclassical limit.

The author would like to thank M. D. Coutinho-Filho for reading the manuscript
and helpful comments. This research was partially supported by the Brazilian
Agency CNPq.

\end{document}